\documentclass[twocolumn,10pt,pre,superscriptaddress,showpacs]{revtex4}

\usepackage{graphicx} %eps figures can be used instead
\usepackage{amsmath}
\usepackage{amsfonts}
\usepackage{amssymb}
\usepackage{amsthm}
\usepackage{array}

\usepackage{color}

\def\longrharpup{\relbar\joinrel\rightharpoonup}
\def\longlharpdn{\leftharpoondown\joinrel\relbar}
\def\rlPOON#1{\vcenter{\hbox{\ooalign{\raise2.3pt
        \hbox{$#1\longrharpup$}\crcr $#1\longlharpdn$}}}}
\def\eqbm{\mathrel{\mathpalette\rlPOON{}}\rm}
\def\eqbmlab#1_#2{\mathrel{\mathop{\eqbm}\limits~{#1}_{#2}}\rm}

\def\yieldslab~#1{\mathrel{\mathop{\longrightarrow}\limits~{#1}} \rm}

\begin{document}

\title{Spatial correlations in nonequilibrium reaction-diffusion problems\\by the Gillespie algorithm}
\author{Jorge Luis Hita}
\author{Jos\'e Mar\'ia Ortiz de Z\'arate}\thanks{Electronic address: jmortizz@fis.ucm.es}
\affiliation{Departamento de F\'{\i}sica Aplicada I. Universidad Complutense. Madrid, Spain}

\begin{abstract}
We present a study of the spatial correlation functions of a one-dimensional reaction-diffusion system in both equilibrium and out of equilibrium. For the numerical simulations we have employed the Gillespie algorithm dividing the system in cells to treat diffusion as a chemical process between adjacent cells. We find that the spatial correlations are spatially short ranged in equilibrium but become long ranged in nonequilibrium. These results are in good agreement with theoretical predictions from fluctuating hydrodynamics for a one-dimensional system and periodic boundary conditions.
\end{abstract}

\pacs{82.20.Wt, 05.40.-a, 82.40.Ck}

\maketitle

\section{Introduction}

Thermal fluctuations in nonequilibrium thermodynamic systems have been widely studied during the last years. One of the most striking features is that for nonequilibrium systems the spatial correlations are generically long-ranged; while thermal fluctuations around equilibrium states are, except in the vicinity of critical points, spatially short-ranged~\cite{DorfmanKirkpatrickSengers}. This can be readily concluded theoretically from fluctuating hydrodynamics~\cite{LandauLifshitz}, \emph{i.e.}, by the use of Langevin equations to describe the spatio-temporal evolution of the fluctuations of the thermodynamic fields~\cite{BOOK}. The nonequilibrium problems most thoughtfully studied theoretically have been quiescent fluids subjected to a temperature gradient, both simple~\cite{KirkpatrickEtAl,RonisProcaccia,Physica2} and binary mixtures~\cite{LawNieuwoudt,SegreSengers}, as well as fluids under flow~\cite{BenaEtAl,LD02,WS03,miORR}. The most salient aspects of the theory have been confirmed experimentally, in particular for the case of fluids subjected to temperature gradients~\cite{LawGammonSengers,VailatiGiglio1,TakacsEtAl2}.

Nonequilibrium fluctuations in reaction-diffusion problems have also been considered theoretically, whether the system is driven outside of equilibrium by a temperature gradient~\cite{miJCP2,miPCCP}, or by an external flow of particles~\cite{NicolisMalekMansour} or by assuming different direct and inverse reaction mechanisms~\cite{BOOK}.

Although some authors have presented numerical simulations of nonequilibrium correlations~\cite{ZhangFan,Gorecki1994,Wakou2003}, the number of investigations is still scarce. Our goal here is to contribute to fill this gap by presenting numerical simulations of equilibrium and nonequilibrium reaction-diffusion problems, for which the Gillespie~\cite{Gillespie1976,Gillespie1977} algorithm is particularly adequate. The Gillespie algorithm is a Monte Carlo method that can simulate efficiently a large network of coupled chemical reactions based on the chemical master equation~\cite{Gillespie1976,Gillespie1977}. Although it was developed in 1976, the interest in this algorithm has grown during the last years, particularly because of its proven utility in biophysical problems, like gene expression~\cite{Gene1}, regulatory networks~\cite{RegNet}, or others~\cite{BarkaiLeibler,WhiteEtAl}. The Gillespie algorithm was originally conceived to describe chemical kinetics, but it can be applied to diffusion processes as well~\cite{Bernstein2005}.

The Gillespie algorithm simulates the chemical master equation exactly, but for real applications in biology and biophysics it is often too slow. One way to speed it up is to solve the master equation approximately, averaging the most common reactions and focusing only in the not-very-frequent ones, as in the tau-leap method~\cite{Gillespie2} or alternative improved approaches based on related ideas~\cite{Gillespie2003}. Recently, a way to accelerate the algorithm while keeping it exact was devised~\cite{Sagar2010}, this method is based on data structures called binary trees. In this approach the updating of every reagent in the system is not needed, but only of the reagents that are involved in the current reaction. Hence, it is possible in some applications to convert the O($n$) original Gillespie algorithm into a O(log~$n$) algorithm~\cite{Sagar2010}. We have not find necessary to use these improvements in our case, because the gain is not very significative.

There are several previous works on numerical simulations of spatial correlation functions in reaction-diffusion problems. For instance Wakou et al.~\cite{Wakou2003} analyzed the spatial correlation function of density fluctuations in a model chemical reaction assuming that the monomolecular decay rate is fast enough to consider the motion of the chemicals as ballistic instead of diffusive. They obtained analytical expressions for the spatial correlation functions under this approximation, and simulated them with molecular dynamics. Gorecki et al.~\cite{Gorecki1994}, in a work closer to our present aims, compared spatial correlations between equilibrium and nonequilibrium (broken detailed balance) systems. Results of molecular dynamics simulations were compared with a theory based on a master equation.

Before closing the introduction we should mention that spatial correlations in nonequilibrium reaction-diffusion systems have been also studied by kinetic theory~\cite{Wakou2000}, an approach that goes beyond fluctuating hydrodynamics and in which the local equilibrium assumption is not needed. These results~\cite{Wakou2000} agree with fluctuating hydrodynamics for sufficiently slow chemical reactions, but also show that for very fast chemical reactions fluctuating hydrodynamics does not apply. Again, in our simulations the hydrodynamic approximation holds and the kinetic theory approach is not very relevant.

The paper is organized as follows. First, in Sect.~\ref{S1}, we describe the two chemical reactions to be studied, one displaying a stationary state that is of equilibrium. The second chemical reaction exhibits broken detailed balance, and the corresponding steady state is out of equilibrium. In Sect.~\ref{S2} we present the theory and obtain analytic results for both the equilibrium and the nonequilibrium processes in the continuous limit. Then, in Sect.~\ref{'simu'} we describe the details of the simulations. In Sect.~\ref{S5} we present the results obtained from the simulations in three steps: applicability of the chemical Langevin equation, spatial correlation function of fluctuations in the number of particles per cell, and a third subsection devoted to details in the comparison with the continuous theory of Sect.~\ref{S2}. Finally, in Sect.~\ref{S6} our main conclusions are summarized.

\section{Equilibrium and nonequilibrium chemical reactions\label{S1}}

Our goal is to obtain from simulations the spatial correlation function of concentration fluctuations in a reaction-diffusion system in equilibrium and in a reaction-diffusion system exhibiting a steady state solution that is out of equilibrium. Several mechanisms have been devised and considered in the literature to drive a reaction-diffusion system out of equilibrium. One possibility is to externally impose a flow, by adding and removing particles~\cite{NicolisEtAl,BarasEtAl2,MareschalDeWit,Wakou2000,Wakou2003}. Another option, that keeps the system closed, is to consider a complex chain of chemical reactions that effectively break detailed balance, like the Schl\"ogl kinetic mechanism~\cite{NicolisMalekMansour} or other variants~\cite{Gorecki1994,Wakou2000,Wakou2003}. A third possibility is that of a single closed chemical reaction in the presence of a temperature gradient~\cite{miJCP2,miPCCP}. For our present purposes we tried to be as simple as possible, so we adopted the kinetic mechanism proposed by van Wijland, Oerding and Hilhorst~\cite{WijlandOerdingHilhorst} (WOH), that consist of a closed isothermal system with only two active chemical species with broken detailed balance due to an inverse chemical reaction mechanism different from the direct one~\cite{WijlandOerdingHilhorst}.

Therefore, in our simulations we shall consider as a representative example of reaction-diffusion at thermodynamic equilibrium a closed system of two active chemical species, $A$ and $B$, where a dissociation-association chemical reaction occurs:
\begin{align}\label{Req}
A & \stackrel{k_{1}}{\longrharpup}~B,&
B & \stackrel{k_{2}}{\longrharpup}~A.
\end{align}
In addition, we consider the system to be spatially extended (one-dimensional), so that there is diffusion of the two species in a passive solvent. For simplicity we assume the two main diffusion coefficients to be identical, $D$, and neglect cross-diffusion effects\footnote{To adopt a diffusion matrix only adds mathematical complications, the physical behavior being qualitatively the same, \emph{i.e.}, short-range spatial correlations in equilibrium~\cite{BOOK}.}, as expected for an almost ideal mixture~\cite{TaylorKrishna}. For the simulation purposes, we consider the extension $L$ of the system divided in $W$ identical cells of size $\Delta{x}=L/W$. In Eq.~\eqref{Req}, $k_1$ and $k_2$ represent the rate constants, \emph{i.e.}, in the $i$ cell the direct reaction proceeds at a rate $k_1 a_i$ (with $a_i$ the concentration of $A$ molecules in cell $i$, $a_i=N_{A,i}/\Delta{x}$) while the inverse reaction proceeds at a rate $k_2 b_i$ (with $b_i$ the concentration of $B$ molecules in cell $i$, $b_i=N_{B,i}/\Delta{x}$). In a closed system where the chemical reactions~\eqref{Req} occur the concentrations shall evolve to homogeneous stationary values, $a_\text{eq}$ and $b_\text{eq}$ given by:
\begin{align}\label{EQ}
a_\text{eq}&=\frac{k_{2}n_{p}}{k_{1}+k_{2}}, &b_\text{eq}&=\frac{k_{1}n_{p}}{k_{1}+k_{2}},
\end{align}
that is an equilibrium state (chemical potentials are equal). In Eq.~\eqref{EQ}, $n_{p}$ represents the total concentration of molecules ($A$ plus $B$), that is an homogeneous quantity under diffusion. Our first goal will be to investigate thermal fluctuations in the number of molecules per cell around the equilibrium concentrations given by~\eqref{EQ}.

We adopt as a representative example of a reaction-diffusion out of thermodynamic a closed system with two active chemical species, $A$ and $B$, evolving in accordance to the WOH kinetic mechanism:
\begin{align}\label{Rneq}
A & \stackrel{k_{1}}{\longrharpup}~B, &
A+B & \stackrel{k_{2}^\prime}{\longrharpup}~2A.
\end{align}
Diffusion will be treated as in the equilibrium case: the two chemicals diffuse in a solvent, with equal main diffusivities $D$ and zero cross diffusion. In Eq.~\eqref{Rneq}, $k_1$ and $k_2^\prime$ represent the rate constants. In the $i$ cell the direct reaction proceeds at a rate $k_1 a_i$, that for the present work we assume to be the same as in the equilibrium case of Eq.~\eqref{Req}. The inverse reaction, that is different from the case of Eq.~\eqref{Req}, proceeds at a rate $k_2^\prime a_i b_i$. In a closed system where the chemical reactions~\eqref{Rneq} occur, the concentrations shall evolve to one of two possible homogeneous steady states~\cite{WijlandOerdingHilhorst}. The first being characterized by $a_i=0$. Of course, fluctuations around this state cannot be gaussian, and that is precisely the reason why this kinetic mechanism has received considerable attention in the literature, being one of the simplest examples of non-linear (or non-gaussian) fluctuations for whose study renormalization or other complicated mathematical techniques are required~\cite{WijlandOerdingHilhorst}. But we are interested here in the second possible steady state, that is characterized by homogeneous and stationary concentrations, $a_\text{ss}$ and $b_\text{ss}$, of $A$ and $B$ molecules given by:
\begin{align}\label{E04}
a_\text{ss}&=n_{p}-\frac{k_{1}}{k_{2}^\prime},&b_\text{ss}&=\frac{k_{1}}{k_{2}^\prime},
\end{align}
that indeed represent a nonequilibrium state since detailed balance is broken. Our goal is to simulate fluctuations around this steady state, compute their spatial correlation, and compare with the equilibrium case. For the stability of the steady state solution~\eqref{E04} we refer to the original publications~\cite{WijlandOerdingHilhorst}, we just mention that simulations were performed at parameter values for which the solution~\eqref{E04} is indeed stable.

One can see, comparing Eqs.~\eqref{Req} and~\eqref{Rneq}, that the WOH kinetics amounts to imagine that the inverse reaction occurs through a mechanism different from that of the direct reaction. This and similar schemes have been criticized on grounds that thermodynamics requires that any chemical reaction must proceed in either of the two directions (from reactants to products and vice versa). Of course, that is correct. However, our purposes here are mostly to illustrate a generic physical principle by a particular example based on numerical simulations. For that goal we find it convenient to keep the problem as simple as possible while maintaining the key ingredient, \emph{i.e.}, breaking of detailed balance. Furthermore, one can think of Eq.~\eqref{Rneq} as a limiting case, when the inverse rates go to zero of a more complicated kinetics, kind of the ones adopted by other investigators~\cite{NicolisMalekMansour}. Not to mention of many literature examples adopting similar one-directional kinetics~\cite{WijlandOerdingHilhorst,Wakou2000,Wakou2003}

\section{Spatial correlation of concentration fluctuations\label{S2}}

A theoretical calculation of the spatial correlation function of concentration fluctuations for the two chemical kinetic mechanisms, Eqs.~\eqref{Req}-\eqref{Rneq}, is presented in section~11.4 of the book by Ortiz de Z\'arate and Sengers~\cite{BOOK}. However, those results refer to an infinite system and are not directly applicable to the simulations to be presented here. Nonequilibrium fluctuations are spatially long-ranged and, hence, strongly depend on the boundary conditions for the fluctuating fields. In this section we present a theoretical derivation of the spatial correlation function for periodic boundary conditions, which are the ones actually used in the simulations.

Our present derivation is based on the chemical Langevin equation. There has been some debate in the literature on whether the use of a Langevin equation is correct for chemical reactions, or whether fluctuations in chemical reactions must always be described by a
chemical master equation~\cite{Gillespie,Gillespie2,Zwanzig}. From a strictly microscopic point of view, we acknowledge that the proper theoretical framework to describe fluctuations in chemical reactions has to be based on the chemical master
equation~\cite{Gardiner}, in particular when a small number of molecules is involved. However, starting from the chemical master equation, through a Kramers-Moyal approximation combined with the system-size expansion proposed by van Kampen~\cite{vanKampen2,vanKampen}, it is possible to obtain a Fokker-Planck (or Langevin) equation which is approximately equivalent to the original chemical master equation~\cite{Gardiner,BarasEtAl}. It is known that this approximation scheme fails when there is a bistability in a system of chemical reactions~\cite{BarasEtAl}, so that the Langevin equation is only valid when fluctuations decay to a single stable solution of the deterministic kinetic equations, and when there are many
particles per unit volume in the system. When presenting our simulations in Sect.~\ref{'simu'}, as a preliminary step, we shall first discuss how many particles per spatial cell are required to justify the use of a chemical Langevin equation to interpret the results.

\subsection{Equilibrium concentration fluctuations\label{S31}}

We first evaluate the spatial correlation of concentration fluctuations in the equilibrium reaction diffusion problem of Eq.~\eqref{Req}. As anticipated, following Gardiner~\cite{Gardiner} and others~\cite{vanKampen2,vanKampen}, we adopt the simplest procedure to study the spatiotemporal evolution of these fluctuations, that is the hydrodynamic approximation given by the chemical Langevin equation. Hence, the evolution of the concentration fluctuations around the equilibrium state~\eqref{EQ} will be described by the following set of linear stochastic partial differential equations~\cite{BOOK}:
\begin{equation}\label{EQF}\begin{split}
\frac{\partial \delta a}{\partial t}&= D~\partial_x^2 (\delta a) - k_1~\delta a + k_2~\delta b - \partial_x\delta{J}^{(A)} + \delta \xi,\\
\frac{\partial \delta b}{\partial t}&= D~\partial_x^2 (\delta b) + k_1~\delta a - k_2~\delta b - \partial_x\delta{J}^{(B)} - \delta \xi,
\end{split}\end{equation}
where $x$ is the only continuous space variable in our 1D problem, while $\delta a(x,t) = a(x,t) - a_\text{eq}$ and $\delta b(x,t) = b(x,t) - b_{eq}$ represent the concentration fluctuations around the equilibrium value~\eqref{EQ}. Notice that here the diffusion fluxes are given in terms of molecules (per surface unit and unit time), while other developments~\cite{BOOK} use diffusion fluxes in terms of mass.

In Eqs.~\eqref{EQF}, following the general rules of fluctuating hydrodynamics~\cite{LandauLifshitz,BOOK}, we have introduced three random dissipative fluxes as thermal forcing terms: $\delta{J}^{(A)}(x,t)$ and $\delta{J}^{(B)}(x,t)$ represent the two independent random diffusive fluxes, while $\delta\xi(x,t)$ represents the random reaction rate~\cite{Gardiner}. The statistical properties of these random dissipative fluxes are specified by the so-called fluctuation-dissipation theorem that in this case reads~\cite{BOOK,Gardiner}:
\begin{equation}\label{FDT}\hspace*{-10pt}\begin{split}
\hspace*{-10pt}&\langle\delta{J}^{(A)}(x,t)~\delta{J}^{(A)}(x^\prime,t^\prime)\rangle=2Da_\text{eq}~\delta(x-x^\prime)~\delta(t-t^\prime),\\
\hspace*{-10pt}&\langle\delta{J}^{(B)}(x,t)~\delta{J}^{(B)}(x^\prime,t^\prime)\rangle=2Db_\text{eq}~\delta(x-x^\prime)~\delta(t-t^\prime),\\
\hspace*{-10pt}&\langle\delta\xi(x,t)~\delta\xi(x^\prime,t^\prime)\rangle=[k_1a_\text{eq}+k_2b_\text{eq}]~\delta(x-x^\prime)~\delta(t-t^\prime),
\end{split}\raisetag{20pt}\end{equation}
while all the cross-correlations are zero. Since in the present case we are considering one-dimensional space, diffusion fluxes are scalar. However, we are not considering coupling between the random  diffusion and reaction rate, as in the more general 3D case. To obtain the third of Eqs.~\eqref{FDT} the van~Kampen system-size expansion~\cite{vanKampen2,vanKampen} was used. One identifies in the noise strength the typical result (forward rate plus backward rate) obtained from such a development~\cite{vanKampen,Gardiner}.

The goal is to obtain the statistical properties of the fluctuating fields from the statistical properties of the random dissipative fluxes~\eqref{FDT}. First, as usual~\cite{SchmitzCohen1,BOOK}, since fluctuations are around a stationary (in this case equilibrium) state, we apply to Eqs.~\eqref{EQF} a Fourier transform in time. Next, to account for periodic boundary conditions in space for the (Fourier transformed) fluctuating fields $\delta{a}(\omega,x)$ and $\delta{b}(\omega,x)$, we express them as:
\begin{equation}\label{E07}
\begin{pmatrix}
\delta{a}(\omega,x)\\\delta{b}(\omega,x)\end{pmatrix}=\sum_{N=-\infty}^\infty \begin{pmatrix}
A_N(\omega)\\B_N(\omega)\end{pmatrix} \exp\left(\mathrm{i}\frac{2N\pi}{L}x\right),
\end{equation}
where $\omega$ is the frequency of the fluctuations and $L$ the spatial periodicity.

By substitution of Eq.~\eqref{E07} into the (Fourier transformed) Eqs.~\eqref{EQF}, after projection onto $(1/L)\exp[-\mathrm{i}(2M\pi/L)x]$, one can readily solve for the amplitudes, namely:
\begin{equation}\label{E08}
\begin{bmatrix}
A_N(\omega)\\B_N(\omega)\end{bmatrix}=\mathsf{M}_N(\omega)
\begin{bmatrix}F_{1,N}(\omega)\\F_{2,N}(\omega)\end{bmatrix},
\end{equation}
where the matrix $\mathsf{M}_N(\omega)$ is given by
\begin{equation}\label{E08B}
\mathsf{M}_N(\omega)=\frac{\begin{bmatrix}
\mathrm{i}\omega+\dfrac{4\pi^2N^2}{L^2}D+k_2 & k_2\\[8pt]k_1 & \mathrm{i}\omega+\dfrac{4\pi^2N^2}{L^2} D+ k_1\end{bmatrix}}
{\left[\mathrm{i}\omega+\dfrac{4\pi^2N^2}{L^2}D\right] \left[\mathrm{i}\omega+D\dfrac{4\pi^2N^2}{L^2}+k_1+k_2\right]},
\end{equation}
and the vector of projected random forces is:
\begin{multline}\label{E09}
\hspace*{-10pt}\begin{bmatrix}F_{1,N}(\omega)\\F_{2,N}(\omega)\end{bmatrix}=\frac{1}{L} \int_0^Ldx~\exp\left(-\mathrm{i}\frac{2N\pi}{L}x\right)\\
\times\begin{bmatrix}-\partial_x\delta{J}^{(A)}(\omega,x) + \delta \xi(\omega,x)\\[12pt]-\partial_x\delta{J}^{(B)}(\omega,x) - \delta \xi(\omega,x)\end{bmatrix},
\end{multline}
with $\delta \xi(\omega,x)$ (and so on) indicating the temporal Fourier transform of the respective random force.

For the evaluation of the correlation function of interest, $\langle\delta{a}(x,t)~\delta{a}(x^\prime,t)\rangle$, we need the correlations between the projected noise terms defined in Eq.~\eqref{E09}. They can be evaluated on the basis of the fluctuation-dissipation theorem, Eq.~\eqref{FDT}, with the result~\cite{BOOK}:
\begin{equation}\label{E10}
\langle{F}_{\alpha,N}^*(\omega)~{F}_{\beta,M}(\omega^\prime)\rangle=C_{\alpha\beta}(N)~\frac{1}{L}\delta_{NM}~2\pi~\delta(\omega-\omega^\prime),
\end{equation}
where we introduce the correlation matrix:
\begin{equation}\label{E12}
\hspace*{-10pt}\mathsf{C}(N)=\begin{bmatrix}
\frac{8N^2\pi^2}{L^2} D a_\text{eq} + \xi^2 & -\xi^2\\
-\xi^2 & \frac{8N^2\pi^2}{L^2} D b_\text{eq}+\xi^2
\end{bmatrix},
\end{equation}
with $\xi^2$ the strength of the random reaction rate in the fluctuation-dissipation theorem of Eq.~\eqref{FDT}, namely
\begin{equation}
\xi^2=k_1 a_\text{eq}+ k_2 b_\text{eq}.
\end{equation}
Next, we proceed to the calculation of the correlation function. In view of Eq.~\eqref{E10} we conclude that it will be expressed as:
\begin{equation}\label{E13}
\langle\delta{a}^*(\omega,x)~\delta{a}(\omega^\prime,x^\prime)\rangle = S(\omega,x,x^\prime)~2\pi~\delta(\omega-\omega^\prime),
\end{equation}
where an explicit expression for $S(\omega,x,x^\prime)$ as a series of exponentials can be readily obtained from Eqs.~\eqref{E07}-\eqref{E12}. We shall not display it here, because it is a long expression not really very informative. Note that for the (equal-time) spatial correlation we only need its frequency integration. Indeed, applying a double inverse Fourier transform to Eq.~\eqref{E13} one readily obtains:
\begin{equation}
\langle\delta{a}(x,t)~\delta{a}(x^\prime,t)\rangle = \frac{1}{2\pi}\int_{-\infty}^\infty d\omega~S(\omega,x,x^\prime),
\end{equation}
indicating that the spatial correlation of concentration fluctuations is independent of the time $t$ at which it is evaluated, as expected for a time-translational invariant equilibrium state.

It turns out that the frequency integral of the quantity $S(\omega,x,x^\prime)$ introduced in Eq.~\eqref{E13} has a very compact expression:
\begin{equation}\label{E15}
\hspace*{-20pt}\begin{split}
\langle\delta{a}(x,t)~\delta{a}(x^\prime,t)\rangle & = a_\text{eq} \frac{1}{L} \sum_{N=-\infty}^\infty \exp\left(\mathrm{i}\frac{2N\pi}{L}(x-x^\prime)\right)\\
& =a_\text{eq}~\delta(x-x^\prime),\raisetag{12pt}
\end{split}
\end{equation}
where the Fourier series expansion of the delta function for periodic boundary conditions has been used.

Therefore, we conclude that the spatial correlation for the equilibrium reaction-diffusion problem is spatially short-ranged. In the hydrodynamic continuous space description we are employing here, $\langle\delta{a}(x,t)~\delta{a}(x^\prime,t)\rangle$ is proportional to a delta function. The proportionality constant equals the equilibrium concentration. Indeed, as is well-known, concentration fluctuations for a chemically reacting system in equilibrium (detailed balance) follow a gaussian distribution whose variance equals the square root of the mean value~\cite{Gardiner}. Of course, this is only possible when the average number of particles per cell in equilibrium is large enough, the same condition that is required to apply a van~Kampen system-size expansion~\cite{vanKampen2,vanKampen} and, thus, to justify the chemical Langevin equation itself.

\subsection{Nonequilibrium concentration fluctuations}

We now evaluate the spatial correlation of concentration fluctuations around the steady state~\eqref{E04} of the nonequilibrium WOH chemical kinetics of Eq.~\eqref{Rneq}. We follow exactly the same scheme as in Sect.~\ref{S31} and continue to adopt the chemical Langevin equation as the starting point. In the nonequilibrium case, the spatiotemporal evolution of the fluctuations around the steady state~\eqref{E04} will be described by the following set of linear stochastic partial differential equations~\cite{BOOK}:
\begin{equation}\label{E16}\begin{split}
\frac{\partial \delta a}{\partial t}&= D~\partial_x^2 (\delta a) + k_2^\prime{a}_\text{ss}~\delta b - \partial_x\delta{J}^{(A)} + \delta \xi,\\
\frac{\partial \delta b}{\partial t}&= D~\partial_x^2 (\delta b) - k_2^\prime{a}_\text{ss}~\delta b - \partial_x\delta{J}^{(B)} - \delta \xi,
\end{split}\end{equation}
where now $\delta a(x,t) = a(x,t) - a_\text{ss}$ and $\delta b(x,t) = b(x,t) - b_\text{ss}$ represent the concentration fluctuations around the nonequilibrium steady state~\eqref{E04}. We have neglected in Eqs.~\eqref{E16} terms quadratic in the fluctuations.

As was the case for the equilibrium fluctuations, Eqs.~\eqref{EQF}, we have considered tree independent sources of stochastic forcing: two random diffusive fluxes and a random reaction rate. The statistical properties of the random diffusion fluxes are the same as in the equilibrium case, Eqs.~\eqref{FDT}. However, the strength of the random reaction rate is different because of the different underlying chemical kinetics. The reaction rate correlation can be obtained by the usual system-size expansion~\cite{Gardiner} for deriving the chemical Langevin equation from the master equation, with the result~\cite{BOOK}:
\begin{multline}\label{FDT2}
\langle\delta\xi(x,t)~\delta\xi(x^\prime,t^\prime)\rangle=[k_1a_\text{ss}+k_2^\prime{a}_\text{ss}b_\text{ss}]\\
\hspace*{30pt}\times~\delta(x-x^\prime)~\delta(t-t^\prime).
\end{multline}
Again, we observe the typical result: rate of the forward reaction plus rate of the backward reaction~\cite{vanKampen,Gardiner}.

Next, the calculation accounting for periodic boundary conditions follows the same steps as in Sect.~\ref{S31} for the equilibrium kinetics. Of course, the matrices $\mathsf{M}_N(\omega)$ in Eq.~\eqref{E08B} and $\mathsf{C}(N)$ in Eq.~\eqref{E12} will be different. The respective expressions can be easily obtained, and we are not displaying it here.

We focus then on the final result for the equal-time spatial correlation function, which results in:
\begin{equation}\label{E18}
\langle\delta{a}(x,t)~\delta{a}(x^\prime,t)\rangle = a_\text{ss}~\delta(x-x^\prime)+ S_\text{ne}(x-x^\prime),
\end{equation}
with
\begin{equation}\label{E19}
S_\text{ne}(u)= -\frac{n_p}{L}+\frac{2k_1}{k_2^\prime{L}} \sum_{N=-\infty}^\infty \frac{\exp\left(\mathrm{i}\dfrac{2N\pi}{L}u\right)} {1+\dfrac{8\pi^2 D N^2}{k_2^\prime a_\text{ss}L^2}}.
\end{equation}
We note in Eq.~\eqref{E18} that the spatial correlation contains two contributions. First is a short-range part (proportional to a delta function) that is the same as if the system were at equilibrium at the steady state concentration, see Eq.~\eqref{E16}. However, because detailed balance is broken in the case of the chemical kinetics~\eqref{Rneq}, there appears an extra nonequilibrium contribution represented by the function $S_\text{ne}(u)$. The sum of the Fourier series in Eq.~\eqref{E19} can be performed analytically~\cite{Gradstein}, giving:
\begin{equation}\label{E20}
S_\text{ne}(u)= -\frac{n_p}{L}+\frac{2k_1\mu}{k_2^\prime{L}\sinh{\mu}} {\cosh{\left[2\mu\left(\dfrac{|u|}{L}-\dfrac{1}{2} \right)\right]}},
\end{equation}
with
\begin{equation}\label{E21}
\mu = \sqrt{a_\text{ss}\frac{k_2^\prime L^2}{8 D}}=\sqrt{\frac{a_\text{ss}k_2^\prime}{8 k_\text{D}}},
\end{equation}
where $k_\text{D}=D/L^2$ is the rate corresponding to diffusion. A simple inspection at Eq.~\eqref{E20} shows that the nonequilibrium contribution to the spatial correlation is long-ranged. Depending on the value of the parameter $\mu$, it can encompass the whole system, as further discussed below.

The results~\eqref{E18}-\eqref{E20} are for periodic boundary conditions, as implemented by Eq.~\eqref{E07}. Therefore they differ from the solution for an infinite (bulk) system presented in Sect.~11.4 of a previous publication~\cite{BOOK}. However, it can be verified that, in the limit $L\to\infty$, the present results converge to the one-dimensional version of Eq.~(11.35) elsewhere~\cite{BOOK}. For comparison with the simulations, Eqs.~\eqref{E18}-\eqref{E20} must be used. Obviously, numerical simulations can only be performed in a system of finite extension, and due to their spatial long-range nature, nonequilibrium fluctuations are strongly affected by the boundary conditions~\cite{BOOK,Physica2}

\section{Simulations}\label{'simu'}

We have performed extensive numerical simulations of the equilibrium and the non equilibrium reaction-diffusion processes introduced in Sect.~\ref{S1} and analyzed theoretically in Sect.~\ref{S2}. For this purpose we used the Gillespie~\cite{Gillespie1976,Gillespie1977} algorithm, that has become very popular lately for simulation of large and complex networks of chemical reactions~\cite{Gene1,RegNet,BarkaiLeibler,WhiteEtAl}. For a detailed description of the algorithm and its many applications, we refer to the original publications~\cite{Gillespie1976,Gillespie1977,Gillespie2003,Sagar2010}

As discussed by Bernstein~\cite{Bernstein2005}, simulation of diffusion processes can be easily integrated into Gillespie algorithm. This integration allows to develop a fast and compact code, where diffusion and chemical reactions are treated on equal footing. The idea of Bernstein~\cite{Bernstein2005} is to divide the space into cells, and to consider molecules in different cells as if they were different chemical species. Diffusion (in one dimension) is included by adding to the list of chemical reactions, for each cell $i$ and each chemical species, two \emph{reactions}; one transforming a molecule in the cell $i$ into a molecule of the same species in the cell $i-1$, and another transforming a molecule in the cell $i$ into a molecule of the same species in the cell $i+1$. This amounts to add a large number of items to the list of chemical reactions, but the Gillespie algorithm is particularly fast and good in handling such large numbers of reactions. It is found that this extra \emph{reactions} associated to diffusion have to proceed at the diffusion rate $k_\text{D}$. Bernstein~\cite{Bernstein2005} has presented a large number of simulations, showing that this method correctly simulates diffusion in one-dimensional systems.

Most of the simulations to be presented in this paper are for a number $W=128$ of cells and periodic boundary conditions, so that they are effectively performed in a ring of 128 cells. This number was selected after some preliminary runs that demonstrated this number of cells enough to show the effects searched for at a reasonable noise level. A smaller number of cells gave a lot of spatial noise, a larger number of cells took a prohibitive long time with the computing means at our disposal. At each site we consider six effective chemical reactions, two corresponding to either the equilibrium~\eqref{Req} or the nonequilibrium~\eqref{Rneq} kinetics, two corresponding to diffusion of $A$ molecules to the two adjacent cells and other two corresponding to diffusion of $B$ molecules to the two adjacent cells. Hence, the total list of chemical reactions comprises $6\times 128 = 768$ items.

The Gillespie algorithm simulates solutions of the master equation, while the theory presented in Sect.~\ref{S2} is based on a  Langevin equation. Therefore, for the simulations to correspond to the theory we need a large enough number of molecules of the two species in all the spatial cells (see discussion at the beginning of Sect.~\ref{S2}). After some preliminary runs, we settled at an average total number of molecules per cell $\langle{N}\rangle=3000$, thus, the number of molecules in the system ($A$ plus $B$) is $128\times3000$, that is a conserved quantity. We selected the reaction rates so that the average number of molecules per cell is the same for the two species, equal to half the total (average) number $\langle{N}\rangle$ of molecules per cell. This selection makes sure that (for a given $\langle{N}\rangle$) the average number of molecules is as far from zero as possible for the two species simultaneously. As discussed later, this is a requirement for the probability distribution of the fluctuations to be as normal (gaussian) as possible. In addition, to set the averages the same for the equilibrium~\eqref{Req} and the nonequilibrium~\eqref{Rneq} kinetics facilitates the comparison of the corresponding spatial correlations. We then select the reaction rate of the first chemical reaction (common to the equilibrium~\eqref{Req} and the nonequilibrium~\eqref{Rneq} kinetics) as $k_1=30$, which is equivalent to fix the time unit of our simulations to the same vale for the two kinetics. Consequently, the rates of the inverse reactions were fixed at $k_2=30$ for the equilibrium kinetics of Eq.~\eqref{Req} and $k_2^\prime=0.02$ for the nonequilibrium WOH kinetics of Eq.~\eqref{Rneq}. These choices set the steady average number of $A$ molecules per cell and the steady average number of $B$ molecules per cell at $\langle{N_A}\rangle=\langle{N_B}\rangle=\langle{N}\rangle/2=1500$ in both cases.

We have performed simulations at various values of the diffusion rate constant  $k_\text{D}$, which we remind is the same for the two species. The space unit is fixed by the length $L$ of the ring, that we take as $L=1$. Hence, the size of the cells that discretize the space is $\Delta{x}=L/W=1/128$. The continuous theory presented in Sect.~\ref{S2} is formulated in terms of concentrations, \emph{i.e.}, molecules per unit volume\footnote{Per unit length in our one-dimensional case}. Thus, the selected number of molecules per cell correspond to steady concentrations $a_\text{eq}=a_\text{ss}=\langle{N_A}\rangle/\Delta{x}=W\langle{N_A}\rangle/L=128\langle{N_A}\rangle$, and similarly for $B$.

All the simulations started from an initial state where $A$ and $B$ molecules where homogeneously distributed among the $W$ spatial cells. Therefore we put 1500 molecules of type $A$ and 1500 molecules of type $B$ in all the 128 cells at $t=0$. Then, we run the algorithm at least $6\times10^9$ iterations for \emph{equilibration}. Only after that we started number tracking for further analysis. Some simulations were performed with an alternative initial state that was setup by randomly selecting an initial cell for each one of the $128\times1500$ $A$ molecules, and the same for the $B$ molecules. We found no difference between the results obtained with the two alternative initial states.

Some runs for the nonequilibrium WOH kinetics of Eq.~\eqref{Rneq} were done with a larger number of cells, $W^\prime=256$, at the expense of a larger computing time. The goal was to simulate the same system but with a finer space resolution. Therefore, the average number of particles per cell was reduced to half the value used for the main runs: $\langle{N^\prime_A}\rangle=750$ and $\langle{N^\prime_B}\rangle=750$, to keep the (average) steady concentrations the same, \emph{i.e.}, the same number of particles per unit length. The time unit $k_1$ was set at the same value as for the main runs. As a consequence, the rate $k^\prime_2$ of the inverse reaction has to be fixed at $k_2^\prime=0.04$ for these runs with 256 cells. Systems with a given diffusion constant $D$ have to be simulated with the same diffusion rate $k_\text{D}$, independent of the number of cells used for a finer description.

For the simulations, we wrote a specific C++ code that was run in a state of the art personal computer. To finalize, we should mention that, for speeding up the process and since for the present work we are only interested in equal-time fluctuations, we often did not keep track of the time, in particular for the simulations used to analyze the spatial correlations.

\section{Results and analysis\label{S5}}

\subsection{Applicability of the chemical Langevin equation\label{S51}}

\begin{figure}[t]
\centering
  \includegraphics[width=\columnwidth]{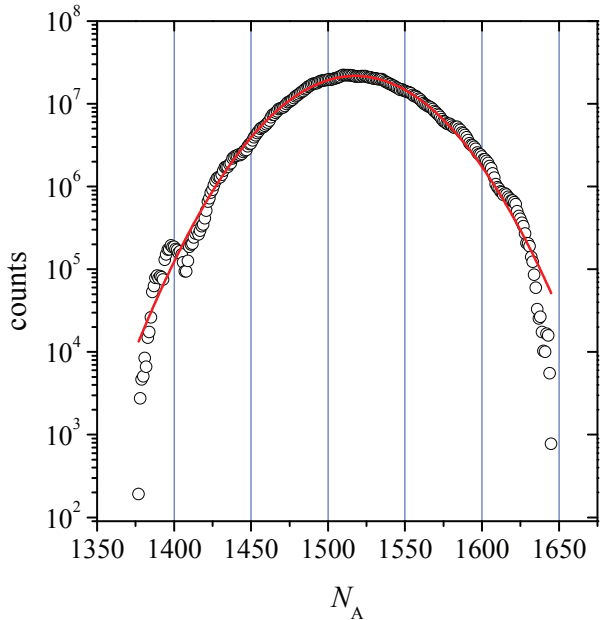}
  \caption{(Color online) Histogram of the number of $A$ molecules in a randomly chosen cell for the equilibrium kinetics, Eq.~\eqref{Req}, with parameter values specified in the text. Number of iterations of the algorithm is $2\times10^9$ for this particular case. The solid line is a fitting to a gaussian. In this particular case $\langle{x}\rangle=1517$ and $\sigma=36.59$}
  \label{fgr:1}
\end{figure}
Before starting an extensive set of simulations, we have to make sure that the chemical Langevin equation used in the theory Sect.~\ref{S2} is adequate to describe the statistical properties of the problem. With such a goal we performed a series of simulations for the equilibrium kinetics of Eq.~\eqref{Req}, tracking the number of $A$ molecules ($N_{A,i}$) in a randomly selected $i$ cell at each time step. Afterwards, we built the corresponding histograms, an example of which is shown in Fig.~\ref{fgr:1}, that corresponds to the cell number 12 and for $k_\text{D}=1$. These histograms were fitted to gaussians, and in all cases the values obtained for the mean $\langle{x}\rangle$ and the variance $\sigma$ were reasonably close to the expected numbers, $\langle{N_A}\rangle=1500$ and $\sigma=\sqrt{1500}\simeq38.73$, on the basis of the chemical Langevin equation of Sect.~\ref{S2}.

\begin{figure}[t]
\centering
  \includegraphics[width=\columnwidth]{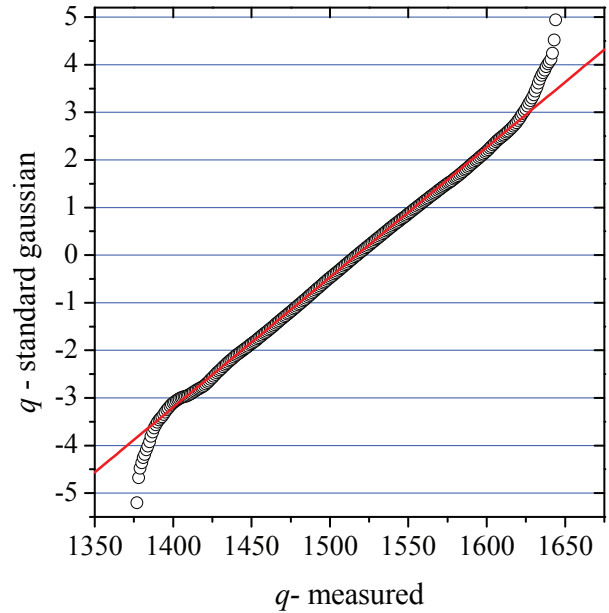}
  \caption{(Color online) Percentiles (q-–q plot) normality test applied to the histogram of Fig.~\ref{fgr:1}. The straight line corresponds to the gaussian fit also shown in Fig.~\ref{fgr:1}. Some deviations from gaussian are observable at the tails, probably due to low sampling}
  \label{fgr:2}
\end{figure}

Trying to be more quantitative, we also performed percentiles normality tests (q–-q plots). In Fig.~\ref{fgr:2} we show an example computed from the same histogram displayed in Fig.~\ref{fgr:1}. In the ordinate we represent the value $x$ calculated from a standard gaussian that accumulates (from $x=-\infty$) the same probability as the percentile accumulated by the corresponding abscissa value in the measured histogram. The solid line ($y=(x-1517)/36.59$ for this particular case) represents the gaussian distribution fitted to  the histogram.

A simple look at Figs.~\ref{fgr:1}-\ref{fgr:2} confirms that the histograms obtained from the simulations are well represented by  normal distributions. Of course, some deviations are observable at the tails, most likely due to low sampling. Note in Fig.~\ref{fgr:2} that significative deviations appear at distances farther than $\pm3\sigma$ from the mean, corresponding to very low probability.

From the data displayed in Figs.~\ref{fgr:1}-\ref{fgr:2} and similar, we conclude that the parameter values selected (in particular the number of particles per cell) are large enough to justify the series of approximations, essentially a system-size expansion~\cite{vanKampen,vanKampen2}, that make the chemical Langevin equation equivalent to the more fundamental master equation. Since the Gillespie algorithm obtains solutions to a master equation, it is particularly well suited for this kind of investigation~\cite{LafuerzaToral}.

We should mention that, before deciding on the current parameter values, we performed simulations with a lower number of particles per spatial cell, and indeed observed deviations from normality. This is somewhat expected when the number of particles per cell $\langle{N_A}\rangle$ is reduced to become a few units of $\sqrt{\langle{N_A}\rangle}$. In that case it is obvious that the probability of a given fluctuation $\delta{N}_A$ cannot be a gaussian. As already mentioned, we settled in our case to a large enough $\langle{N_A}\rangle=1500$ to make  deviations from normality negligible.

The results presented in this subsection provide further evidence of the adequacy of Bernstein~\cite{Bernstein2005} adaptation of the Gillespie~\cite{Gillespie1976,Gillespie1977} algorithm for reliable simulation of reaction-diffusion processes. Bernstein has shown~\cite{Bernstein2005} that this extension of the algorithm give concentration profiles that, on average, evolve according to the diffusion equation. Here, we have shown that the statistical properties of the fluctuations around a steady state also reproduce the expected behavior for diffusion.

\subsection{Spatial correlations}

\begin{figure*}[ht]
\centering
  \includegraphics[width=\textwidth]{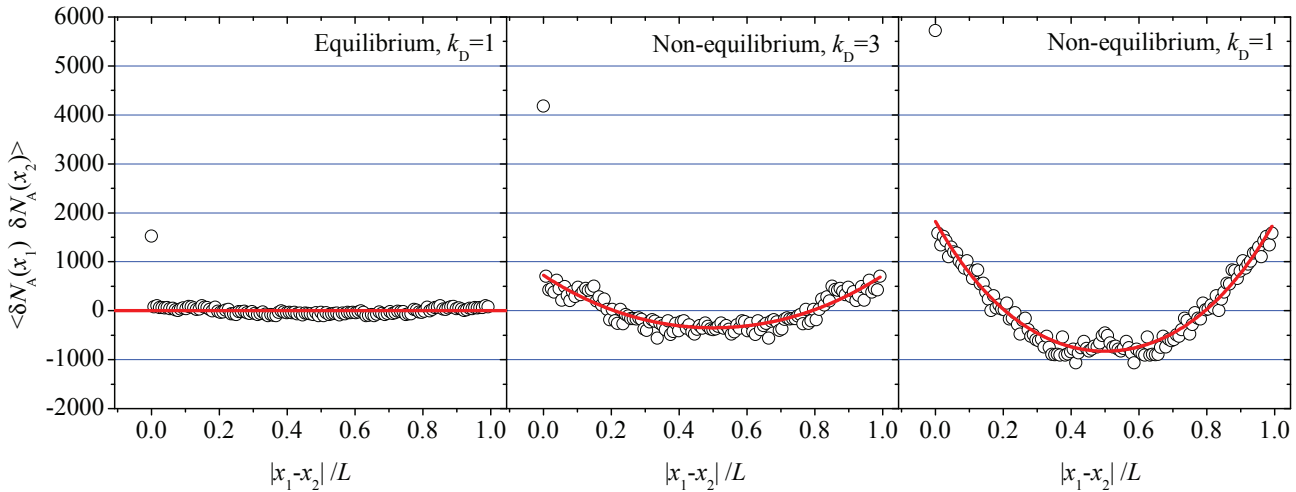}
  \caption{(Color online) Three examples of spatial correlation functions in reaction-diffusion problems. Left panel corresponds to the  equilibrium kinetics of Eq.~\eqref{Req} and center and right panels correspond to the nonequilibrium WOH kinetics of Eq.~\eqref{Rneq}. The $k_\text{D}$ values are indicated. Other parameter values are given in the text. Number of spatial cells is $W=128$ in all cases. For the nonequilibrium panels, the solid curves represent a fitting to Eq.~\eqref{E20} as further discussed in the main text. Note that the vertical scale is the same for the three cases}
  \label{fgr:4}
\end{figure*}

Next, we discuss the spatial correlation function $\langle\delta{N}_\text{A}(x)~\delta{N}_\text{A}(x^\prime)\rangle$ of equal-time fluctuations in the number of $A$-particles per cell. We evaluated this correlation function by direct spatial average, namely:
\begin{equation}\label{E22}
\langle\delta{N}_\text{A}(x)~\delta{N}_\text{A}(x^\prime)\rangle = \frac{1}{W} \sum_{i=1}^W (N_{\text{A},i}-\overline{N}_\text{A}) (N_{\text{A},i+j}-\overline{N}_\text{A}),
\end{equation}
where $N_{\text{A},i}$ is the number of $A$-particles in the cell $i$, and $\overline{N}_\text{A}$ the average number of $A$-particles over all cells. We remind from Sect.~\ref{S2} that for periodic boundary conditions the spatial correlation function depends only on the difference $x-x^\prime=j~\Delta{x}=j~L/W$, a fact that justifies the spatial average conducted in Eq.~\eqref{E22}. Moreover, the finally reported correlations are averaged over various (typically, from six to ten) individual $\langle\delta{N}_\text{A}(x)~\delta{N}_\text{A}(x^\prime)\rangle$ obtained at different times. Hence, we continuously run the algorithm for a large number of iterations, saving the configurations $\{N_{\text{A},i},N_{\text{B},i}\}$ every $2\times10^{9}$ iterations. Later, individual correlation functions were evaluated applying Eq.~\eqref{E22} to these saved configurations. We found it important to wait for a large number of iterations between consecutive saved configurations, because of the long time it takes to dissipate some large fluctuations that appear frequently, in particular for the case of the nonequilibrium WOH kinetics of Eq.~\eqref{Rneq}. Note that in this paper we do not discuss the dynamics of fluctuations that, depending on the value of the diffusion rate $k_\text{D}$, can indeed be very slow.

As an example of the obtained spatial correlation functions, we show in Fig.~\ref{fgr:4} three different cases. Left panel corresponds to the  equilibrium kinetics of Eq.~\eqref{Req} with $k_\text{D}=1$, center panel is for the nonequilibrium WOH kinetics of Eq.~\eqref{Rneq} with $k_\text{D}=3$ and right panel is for the nonequilibrium WOH kinetics of Eq.~\eqref{Rneq} with $k_\text{D}=1$. For easier comparison, the same vertical scale is used to represent the three correlation functions. A simple look at Fig.~\ref{fgr:4} shows our  main result: For equilibrium reaction-diffusion the spatial correlation is zero everywhere, except for $x_1=x_2$, hence it is spatially short-ranged, proportional to a delta function as in Eq.~\eqref{E15}. However, when the reaction-diffusion process is out of (global) equilibrium, the correlation of the fluctuations becomes spatially long-ranged. Actually, for the periodic boundary conditions and parameter values considered here, the range of the correlation encompassed the whole system as it is clearly observed both in the central and right panels of Fig.~\ref{fgr:4}.

Another typical feature of nonequilibrium fluctuations clearly observed in Fig.~\ref{fgr:4} is the intensity enhancement. The leftmost point in each panel of the figure corresponds to fluctuations in the same cell, $\langle\delta{N}_\text{A}^2\rangle$. For the equilibrium case (left panel) we observe $\langle\delta{N}_\text{A}^2\rangle\simeq 1500$, as discussed in more detail in Sect.~\ref{S51}. However, in the center and right panels of Fig.~\ref{fgr:4}, one sees an important enhancement for the nonequilibrium WOH kinetics of Eq.~\eqref{Rneq}. This enhancement increases as the diffusion rate decreases, in accordance to Eq.~\eqref{E20}. Indeed, the curves in the central and right panels of Fig.~\ref{fgr:4} represent fittings of the results to Eq.~\eqref{E20}, using $\mu$ as the only fitting parameter. We leave for the next section a quantitative discussion of these fittings, while we stress here that they can be considered as good. We conclude that our numerical results are well represented by the continuous theory of Sect.~\ref{S2}. In particular, the numerical results contribute to justify one of the approximations adopted in the theory, namely, neglecting nonlinear terms in the fluctuations in Eq.~\eqref{E16}.

Finally, we should mention that in none of the simulations performed for the nonequilibrium WOH kinetics of Eq.~\eqref{Rneq} we observed decay to the other possible homogeneous stationary solution, $a_\text{ss}=0$.
\begin{figure}[th]
\centering
  \includegraphics[width=\columnwidth]{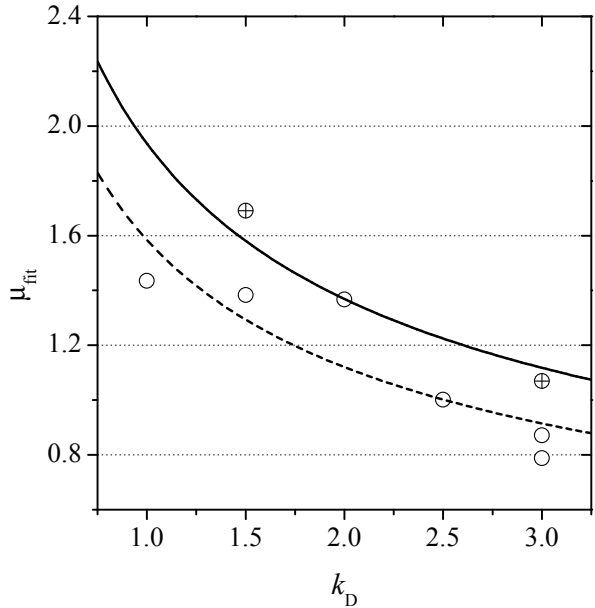}
  \caption{Comparison with the theory of the nonequilibrium spatial correlation functions. The $\mu_\text{fit}$ values obtained from the fits are represented \emph{vs.} the $k_\text{D}$ values used in the corresponding simulations. Open symbols correspond to runs with $W=128$ spatial cells and the $\oplus$ to runs with $W=256$ cells. The solid curve represents the theoretical Eq.~\eqref{E21}. The dotted curve is a fit of the points obtained with $W=128$ cells.}
  \label{fgr:5}
\end{figure}

\subsection{Discussion}

Next, we present a discussion of some quantitative details of our results, that we preferred to separate from the big picture presented in the previous subsection.

Regarding the equilibrium spatial correlations, like the one plotted in the left panel of Fig.~\ref{fgr:4}, they do indeed correspond to delta functions when interpreted as fluctuations in the concentrations and not in the number of particles. This is best understood if one imagines having performed the same simulations but with a larger number of spatial cells. For instance, if the number of cells is doubled (keeping the size of the system constant, $L=1$), the size of the new cells will be $\Delta{x}^\prime=\Delta{x}/2$. In such a case and to simulate the same problem, the average number of particles per cell, $\langle{N_A}\rangle$ and $\langle{N_B}\rangle$, has to be halfed to keep the same concentration in particles per unit length. Hence, a simulation of this equivalent double number of cells system will give a correlation function that is zero everywhere except for the same cell, where will have the value $\langle\delta{N_A^\prime}^2\rangle=\langle\delta{N_A}^2\rangle/2$. To express the spatial correlation function in terms of concentrations one has to divide by the square of the cell size. As a consequence, in terms of concentration fluctuations, one has $\langle\delta{a^\prime}^2\rangle=2 \langle\delta{a}^2\rangle$. We conclude that if one uses a finer spatial grid that is half the original, the height of the only non-zero point in the spatial correlation of the equilibrium concentration fluctuations will double. Hence, in the continuous $\Delta{x}\to 0$ limit the spatial correlation does indeed converge to a delta function.

Regarding the spatial correlations corresponding to the nonequilibrium WOH kinetics of Eq.~\eqref{Rneq}, we tried to explain them on the basis of the continuous theory of Sect.~\ref{S2}. However, as also seen in Fig.~\ref{fgr:4}, the nonequilibrium correlation functions are quite noisy, more than the equilibrium runs at the same $k_\text{D}$. Therefore, we find it best to perform fittings to the theoretical result of Eq.~\eqref{E20}. For these fittings we fix the values of $N_p$ and $k_1/k_2$ to the ones actually used in the simulations ($N_p=3000$ and $k_1/k_2=1500$, for $W=128$ cells), and used $\mu$ as the only fitting parameter. The solid curves displayed in the central and right panels of Fig.~\ref{fgr:4} represent the results of such one-free-parameter fit, that give quite reasonably results. Next, the $\mu$ values obtained from the fits are to be compared with the theoretical value of Eq.~\eqref{E21}.

Such a comparison is shown in Fig.~\ref{fgr:5}, where the $\mu$ values obtained from the fits are represented as a function of the $k_\text{D}$ values used in the corresponding simulations. Simulations performed with $W=128$ cells are represented as open symbols. The solid curve represents the theoretical Eq.~\eqref{E21} for $a_\text{ss}k_2=30$, that is the value used in the simulations. From Fig.~\ref{fgr:5} we first note an important scatter in the $\mu_\text{fit}$, likely associated to the large spatial noise that was observed in the nonequilibrium correlation functions of Fig.~\ref{fgr:4}. It is also noticed that the $\mu_\text{fit}$ values are systematically lower than the theoretical result of Eq.~\eqref{E21}. Indeed, we added as a dashed line a fit to the function $\sqrt{P_1/x}$ (with $P_1$ a fitting parameter) that it clearly lies below the solid line representing the theoretical result. We attribute this difference to the theory of Sect.~\ref{S2} being for continuous space $x$. For confirmation, we run a couple of simulations with a larger number of cells $W=256$ and, consequently, with a smaller average number of particles per cell $N_p=1500$ and larger $k_2^\prime=0.04$ so as to represent the same system (same concentrations $a_\text{ss}$) at increased spatial resolution. We added the $\mu_\text{fit}$ values obtained from these larger scale simulations as $\oplus$ symbols to the data plotted in Fig.~\ref{fgr:5}. We observe that these two additional runs confirm our hypothesis: as finer spatial resolution is used in the simulations, the $\mu$ values obtained from fits approach the theoretical value of Eq.~\eqref{E21}. If one adds up all these clues, it is lead to the conclusion that the simulations confirm the theory presented in Sect.~\ref{S2} in the continuous limit.

\section{Conclusions\label{S6}}

We have presented simulations of two different reaction-diffusion problems using the Gillespie algorithm. One set of simulations represent a chemical reaction in equilibrium; while the other set of simulations correspond to a problem where detailed balance was broken and, consequently, the corresponding steady state is out of equilibrium. From our simulations we obtained and studied histograms of the probability distribution of the number of particles in a given spatial cell, that for the parameter range selected resulted gaussian in good approximation. We also evaluated from the simulations the spatial correlation function of equal-time fluctuations in the number of particles per cell. From this study we have confirmed in a very visual and intuitive way some of the most salient features of nonequilibrium fluctuations. Namely, an intensity enhancement and correlations becomimg spatially long-ranged (as opposed to short-ranged spatial correlations in equilibrium).

The long-range nature of nonequilibrium spatial correlations means that they are affected by boundary conditions. For the periodic boundary conditions used in the simulations, we have developed a continuous theory that gives very compact analytical expressions for the spatial correlations of both the equilibrium and the nonequilibrium problems. The results obtained from the simulations approach the theoretical values in the continuous limit, as the number of spatial cells $W$ is increased.

\emph{Note added in proofs.} Recently a paper by Gillespie et al.~\cite{Gillespie13} appeared, where more background on the application of the Gillespie algorithm to rection-diffusion problems can be found.

\section*{Acknowledgements}

We are indebted to Professor Jan V. Sengers for suggesting part of this research and carefully reviewing the manuscript.

\bibliography{gillespie,ortiz}

\end{document}